# IceCube: Construction Status and First Results

Albrecht Karle, for the IceCube Collaboration

*University of Wisconsin-Madison, 1150 University Avenue, Madison, WI 53706*

**Abstract**

IceCube is a 1 km$^3$ neutrino telescope currently under construction at the South Pole. The detector will consist of 4800 optical sensors deployed at depths between 1450 m and 2450 m in clear Antarctic ice evenly distributed over 80 strings. An air shower array covering a surface area of 1 km$^2$ above the in-ice detector will measure cosmic ray air showers in the energy range from 300 TeV to above 1 EeV.

The detector is designed to detect neutrinos of all flavors: $\nu_e$, $\nu_\mu$ and $\nu_\tau$. With 40 strings currently in operation, construction is 50% complete. Based on data taken to date, the observatory meets its design goals and currently exceeds the sensitivity of AMANDA and previous neutrino telescopes. The construction outlook and possible future extensions are also discussed.

## 1. Introduction

IceCube is a large kilometer scale neutrino telescope currently under construction at the South Pole. With the ability to detect neutrinos of all flavors over a wide energy range from about 100 GeV to beyond $10^9$ GeV, IceCube is able to address fundamental questions in both high energy astrophysics and neutrino physics. One of its main goals is the search for sources of high energy astrophysical neutrinos which provide important clues for understanding the origin of high energy cosmic rays.

The interactions of ultra high energy cosmic rays with radiation fields or matter either at the source or in intergalactic space result in a neutrino flux due to the decays of the produced secondary particles such as pions, kaons and muons. The observed cosmic ray flux sets the scale for the neutrino flux and leads to the prediction of event rates requiring kilometer scale detectors, see for example [1]. As primary candidates for cosmic ray accelerators, AGNs and GRBs are thus also the most promising astrophysical point source candidates of high energy neutrinos.

Galactic source candidates include supernova remnants, microquasars, and pulsars. Guaranteed sources of neutrinos are the cosmogenic high energy neutrino flux from interactions of cosmic rays with the cosmic microwave background and the galactic neutrino flux resulting from galactic cosmic rays interacting with the interstellar medium. Both fluxes are small and their measurement constitutes a great challenge. Other sources of neutrino radiation include dark matter, in the form of supersymmetric or more exotic particles and remnants from various phase transitions in the early universe.

The relation between the cosmic ray flux and the atmospheric neutrino flux is well understood and is based on the standard model of particle physics. The observed diffuse neutrino flux in underground laboratories agrees with Monte Carlo simulations of the primary cosmic ray flux interacting with the Earth's atmosphere and producing a secondary atmospheric neutrino flux [2].

Although atmospheric neutrinos are the primary background in searching for astrophysical neutrinos, they are very useful for two reasons. Atmospheric neutrino physics can be studied up to PeV energies. The measurement of more than 50000 events per year in an energy range from 500 GeV to 500 TeV will make IceCube a unique instrument to make precise comparisons of atmospheric neutrinos with model predictions. At energies beyond 100 TeV a harder neutrino spectrum may emerge which would be a signature of an extraterrestrial flux. Atmospheric neutrinos also give the opportunity to calibrate the detector. The absence of such a calibration beam at higher energies poses a difficult challenge for detectors at energies targeting the cosmogenic neutrino flux.

## 2. Detector

IceCube is designed to detect muons and cascades over a wide energy range. The string spacing was chosen in order to reliably detect and reconstruct muons in the TeV energy range and to precisely calibrate the detector using flashing LEDs and atmospheric muons. The optical properties of the South Pole Ice have been measured with various calibration devices [3] and are used for modeling the detector response to charged particles. Muon reconstruction algorithms [4] allow measuring the direction and energy of tracks from all directions.

In its final configuration, the detector will consist of 80 strings reaching a depth of 2450 m below the surface. There are 60 optical sensors mounted on each string equally spaced between 1450 m and 2450m depth. In addition there will be 320 sensors deployed in 160 IceTop tanks on the surface of the ice directly above the strings. Each sensor consists of a 25 cm photomultiplier tube (PMT), connected to a waveform recording data acquisition circuit capable of resolving pulses with nanosecond precision and having a dynamic range of at least 250 photoelectrons per 10 ns. With the most recent construction season ending in February 2008, half of the IceCube array has been deployed.

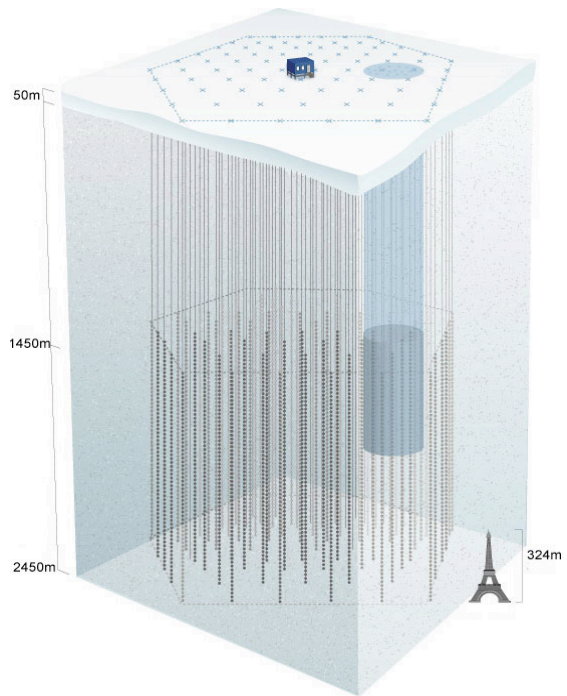

Figure 1: Schematic view of IceCube

Each sensor, hereforth referred to as Digital Optical Module (DOM), collects data autonomously, sending packetized data to the surface. A 33.0 cm diameter pressurized glass sphere holds the Hamamatsu R7081-02 photomultiplier tube plus associated electronics. These electronics include a PMT base, a high voltage generator, a resistive divider PMT base, a flasher board (containing 12 light emitting diodes, with programmable drivers), and a "Main Board" containing a complete data acquisition (DAQ) system [5,6]. The DAQ includes two separate waveform digitizer systems. The first is the analog transient waveform digitizer (ATWD), which uses a custom switched-capacitor array chip to collect 128 samples of the PMT output at 300 megasamples per second (MSPS). The ATWD has three independent channels for each PMT providing 14 bits of dynamic range. The second uses a commercial 40 MSPS 10-bit flash ADC chip, which can record 6.4 μs of data after each trigger.

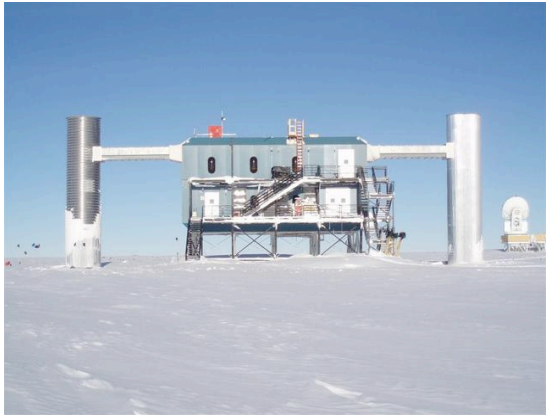

Figure 2: IceCube Laboratory building at the South Pole.

The surface air shower array, IceTop [7], consists of ice Cherenkov detector tanks each containing two DOMs, which are operated at different gain for increased dynamic range. Two such tanks are associated with each string. The tanks are embedded in the snow just below the surface to avoid snowdrift.

The data are collected in a central counting house, located at the geometric center of the IceTop array (Figure 2). The triggering is designed to select a variety of events: muons of all directions, cascades, various topologies of low energy events, and events that are triggered by the AMANDA array (which is operated as a low energy subsystem in IceCube). Once a trigger is issued, hits close to the trigger times are collected and sent to a filtering farm in which the physics streams are selected and the data volume is reduced from about 200 GB/day to about 40 GB/day, a data volume that is small enough to be transmitted by satellite to the data center in the North.

### 3. Construction

The detector is constructed by drilling holes in the ice, one at a time, using a hot water drill. Drilling is immediately followed by deployment of a detector string into the water-filled hole. The drilling of a hole to a depth of 2450 m takes about 35 hours. The subsequent deployment of the string typically takes less than 10 hours. The holes typically freeze back within 1-3 weeks. This method has been pioneered by AMANDA. During the 2007/2008 construction season, the use of a more powerful drill with 5 MW of thermal power combined with a very skilled and motivated crew of drillers and deployers allowed the deployment of 18 strings. The time delay between two subsequent drilling cycles and string deployments was in some cases shorter than 50 hours. By the end of February 2008, 40 strings and 80 IceTop tanks have been deployed.

### 4. Commissioning and Operation

During construction, the detector strings and tanks deployed during the previous seasons are operated at a duty factor of more than 80%. The ability to run the detector during the construction season is based on the relative noise immunity of the detector to electromagnetic interference. During the six week period from February to March after the construction season, the strings are commissioned and the DAQ and filtering programs are finalized in order to prepare for the new configuration. Once the strings are completely frozen in the commissioning can start. Approximately 99% of the deployed DOMs have been successfully commissioned

The 22-string detector configuration (IC22) was in operation from May 2007 to the end of March 2008. Since April 1st, 2008, after the commissioning of the newly deployed strings, the DAQ and the filtering process, IceCube has been running in the 40-string configuration (IC40). The uptime during the first four months has been above 95%. Occasionally, and with decreasing frequency, revised software is released for the DAQ.

### 5. Data

The data consists of events composed of PMT waveforms that were recorded with several waveform recording devices of different gain and sampling rate. The raw digital information is assembled into a calibrated waveform and compressed by a process referred to as feature extraction, where the waveform is described as a superposition of standard pulses with variable amplitude. The comparison of feature-extracted data with raw data shows good agreement and acceptable loss of information.

The DAQ trigger condition is kept for each event and the absolute time is recorded to a precision of about 100 ns. This data is written to tape at the South Pole. A data set for satellite transmission is generated by a filtering process, which reduces the data set by about a factor of 5. There are more than 10 different streams which are generated by special filter requests. Some examples are given for illustration:

- Upgoing muons
- Extremely high energy events
- Moon (for shadow of the moon)
- Cascade like events
- Cosmic ray events
- Ultra low energy events
- WIMPs

These filters are designed by working groups in the collaboration and are reviewed by a trigger and filtering board.

The average filtered data rate of IC22 in 2008 was about 37GB/day. Each data stream is reprocessed in the Northern hemisphere data center, where more computing power is available and more sophisticated reconstruction algorithms can be applied.

## 6. Muons

At the depth of IceCube, the event rate from downgoing atmospheric muons is close to 6 orders of magnitude higher than the event rate from atmospheric neutrinos. In order to avoid contamination of the upgoing neutrino event sample with misreconstructed downgoing events, some quality cuts are required beyond trigger level. With the increasing size of the detector, the probability of randomly coincident muons that can appear as upgoing events for simple reconstruction algorithms increases. Such events may appear as upgoing for simple reconstruction algorithms. An effective separation of the upgoing neutrinos from downgoing background could be achieved with relatively few quality cuts. Figure 3 shows a reconstructed zenith distribution up to right above the horizon compared with simulation. The median angular resolution for muons in IC22 is 1.5°. Figure 4 shows the cumulative point spread function for IC22. The angular resolution obtained with the MonteCarlo simulations can be verified by comparing in-ice events with coincident air shower events or with AMANDA events.

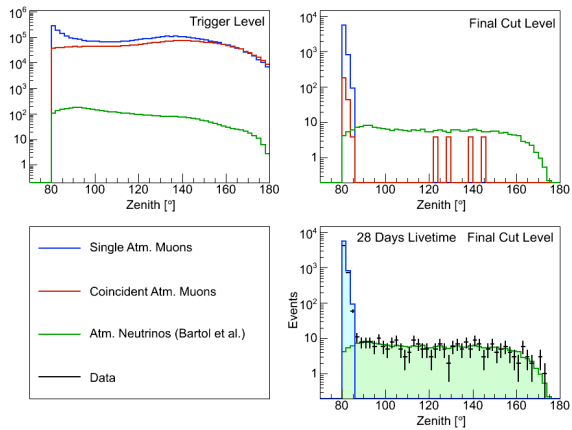

Figure 3: Reconstructed Zenith distribution for data and simulation at trigger level and after the application of quality cuts.

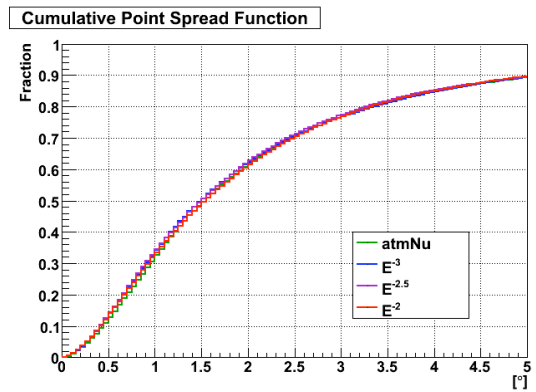

Figure 4: Cumulative Point Spread Function for IC22. The fraction of events contained inside an angle is plotted versus the angular distance from the source.

Another test that is coming quickly within reach of statistics is the observation of the shadowing of cosmic rays by the moon. In a preliminary analysis, the shadow of the moon could be seen with a significance of about 2σ with IC22. In three months of IC40 data 88202 downgoing muon events were observed in a search bin of 1.3° radius where close to 90,000 events were expected, a deficit of 4.2σ. This observation provides a first verification of the absolute pointing capability of IceCube. According to this analysis, the significance of the deficit is within the expectations based on the angular resolution of the instrument.

**7. Search for astrophysical neutrinos**

The search for astrophysical neutrinos can be performed in several ways. Astrophysical neutrinos can be distinguished from atmospheric neutrinos based on:
a) energy → diffuse analysis
b) direction and energy → point source analysis
c) direction, energy and time structure → time dependent point source analysis
d) direction, energy, coincident observation in other observatories → gamma ray bursts, other multi-wavelength analysis

The all sky map of seven years of data taken with the AMANDA detector for the point source analysis is shown in Figure 5.

For IceCube, ten months of the 22-string data (IC22) were recorded and analyzed. At the time of this workshop, the results were not yet unblinded.

The energy response of the IC22 array looks very similar to AMANDA, despite the significantly larger string spacing of 125 m as compared to about 50 m. The much larger scale and the improved angular resolution result in a sensitivity that is almost a factor of 2 better with 0.8 years of IceCube compared to 3.8 years of AMANDA livetime. The robustness and reliability of IceCube data taking also resulted in a livetime of 95% compared to 55% during the seven years of operation of AMANDA. The upper limits derived from AMANDA and the sensitivity curves of different IceCube configurations are shown in Fig. 6.

Figure 7 shows the neutrino effective area for the current point source analysis. The effective area rises strongly until about 1 PeV because of the increasing neutrino nucleon cross section and the increase in the muon range making the detector very effective at higher energies. Above 1 PeV it continues to increase to reach a neutrino effective area of almost 1000 $m^2$ at close to horizontal direction.

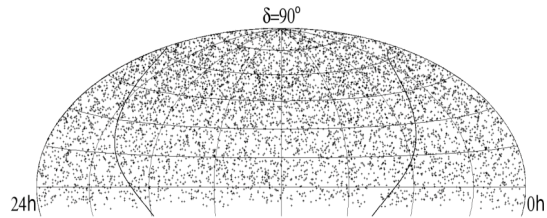

Figure 5: Equatorial sky map of 6595 events recorded by AMANDA from 2000–2006.

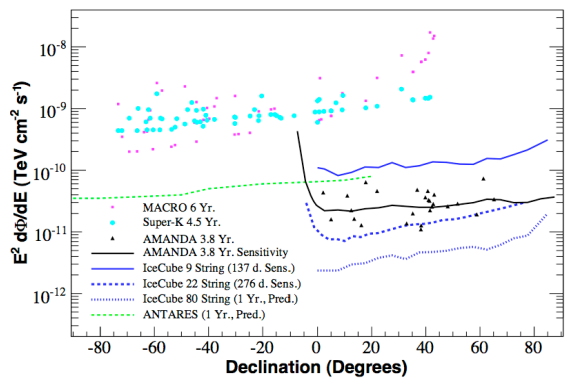

Figure 6: Point source upper limits for AMANDA and IceCube.

Specialized analyses are in development to look for point sources in the region closer to and above the horizon and to look specifically for extremely high

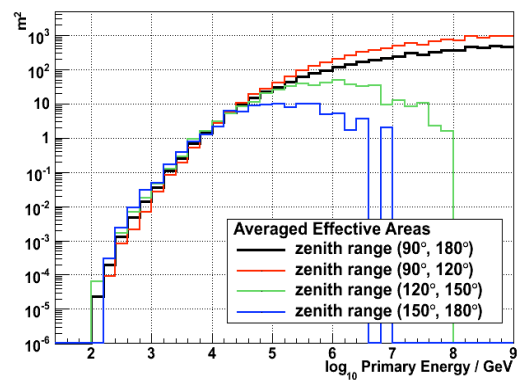

Figure 7: Neutrino effective areas are shown for the IceCube 22 string detector. At large zenith angle one can see the effect of absorption by the Earth.

energy cosmogenic neutrinos.

## 8. Search for dark matter

IceCube is already an effective muon detector at energies slightly above 100 GeV. The search for low energy muons from the direction of the center of the Earth or the Sun can be used for the indirect detection of dark matter. Prominent candidates for dark matter are neutralinos, which may get gravitationally trapped by massive objects and would aggregate in the center of the Earth and the Sun. With an increasing density over cosmic time scales the particles may annihilate with each other and generate a neutrino flux that would be observable with IceCube. The planned extension of IceCube with a Deep Core will substantially enhance the sensitivity of IceCube at energies well below 100 GeV. The predicted neutrino flux from the Sun depends on model assumptions such as the annihilation cross section, neutralino mass and annihilation channel. Our current upper limits on the muon flux from neutralino annihilations obtained with AMANDA and IC22 as well as the projected sensitivity for the full IceCube (IC80) are shown in Figure 8. These limits are compared to other indirect and to direct searches. As can be seen, IceCube with the addition of the Deep Core will improve the reach in the search for dark matter by more than a factor ten compared to AMANDA at all energies.

## 9. IceCube at the highest energies

The capability of IceCube to detect events with energies ranging from 1 PeV to beyond 1 EeV will now be discussed. The two areas of cosmic ray physics and neutrino detection at highest energies will be discussed separately.

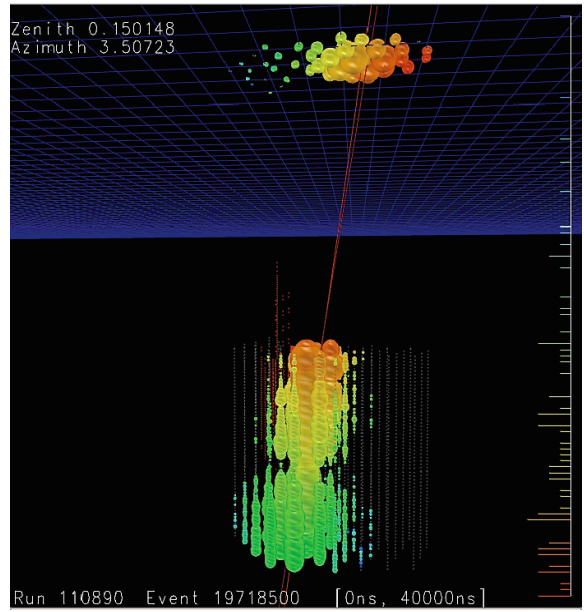

Figure 9: A very high energy cosmic ray air shower observed both with the surface detector IceTop and the in-ice detector string array.

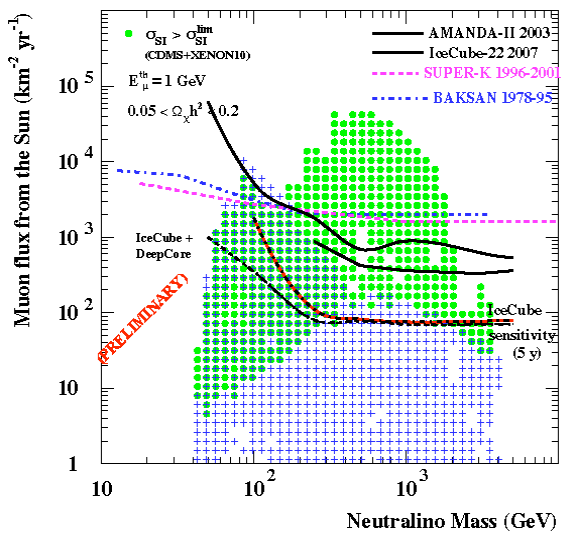

Figure 8: 90% confidence level upper limits on the muon flux from neutralino annihilations in the center of the Sun are compared to other indirect and direct searches.

The surface detector IceTop consists of ice Cherenkov tank pairs. Each IceTop station is associated with an IceCube string. With a station spacing of 125 m, it is efficient for air showers above energies of 1 PeV. Figure 9 shows an event display of a very high energy (~EeV) air shower event. Hits are recorded in all surface detector stations and a large number of DOMs in the deep ice. Based on a preliminary analysis some 2000 high energy muons would have reached the deep detector in this event if the primary was a proton and more if it was a nucleus. With 1 km$^2$ surface area, IceTop will acquire a sufficient number of events in coincidence with the in-ice detector to allow for cosmic ray measurements up to 1 EeV. The directional and calorimetric measurement of the high energy muon component with the in-ice detector and the simultaneous measurement of the electromagnetic particles at the surface with IceTop will enable the investigation of the energy spectrum and the mass composition of cosmic rays.

Events with energies above one PeV can deposit an enormous amount of light in the detector. Figure 10 shows an event that was generated by flasher pulse produced by an array of 12 blue LEDs that are mounted on every IceCube sensor. The event produces an amount of light that is comparable with that of an electron cascade on the order of 1 PeV. Photons were recorded on strings at distances up to 600 m from the flasher. The events are somewhat brighter than previously expected because the deep ice below a depth of 2100m is exceptionally clear. The scattering length is substantially larger than in average ice at the depth of AMANDA.

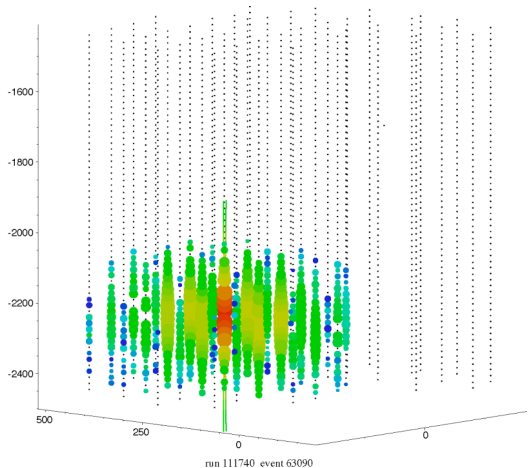

Figure 10: A flasher event in IceCube. Such events, produced by LEDs built in the DOMs, can be used for calibration purposes.

Extremely high energy (EHE) events, above about 1 PeV, are observed near and above the horizon. At these energies, the Earth becomes opaque to neutrinos and one needs to change the search strategy. In an optimized analysis, the neutrino effective area reaches about 4000 m$^2$ for IC80 at 1EeV energy. IC80 can therefore test optimistic models of the cosmogenic neutrino flux. IceCube is already accumulating an exposure with the current data that makes detection of a cosmogenic neutrino event possible. The first EHE analysis is underway and will be published soon.

## 10. Future developments

The IceCube construction schedule plans for three more seasons with the completion of IceCube scheduled in the season 2010/11. In addition to the baseline configuration, the collaboration is pursuing the goal of adding a densely instrumented Deep Core consisting of 6 strings [8]. These strings would be added to the center of the array with most of the sensors (50 out of 60/string) deployed with small spacing in the clear ice below 2100 m. This Deep Core is expected to substantially enhance the sensitivity at low energies.

Another extension in consideration is for high energies. A number of the outer strings may be deployed at larger spacing to enhance the high energy response of the detector (Figure 11). A modified arrangement of the last 10 strings or so could give an increase of the effective area by 20% or more above the baseline at energies above 1 PeV. This modification would significantly enhance the sensitivity of the detector to search for a diffuse astrophysical neutrino flux and the cosmogenic GZK [9] neutrino flux. Only at energies below a few TeV would there be a small reduction in effective area.

It should be noted that the reliable detection of the cosmogenic neutrino flux at the highest energies would require an even larger detector. A recent review of predicted fluxes can be found in [10]. The collaboration is pursuing research and development activities to investigate new techniques of acoustic and radio detection of these highest energy neutrinos. The details of these efforts are being presented in several papers at this workshop. These activities can be grouped in three categories:

- Acoustic detector development: The South Pole ice is expected to have a large attenuation length on the order of 1 km or

more. First results of acoustic sensor R&D and measurements in South Pole ice are shown at this meeting [11]. More measurements are being planned.
- Radio detector development: In the tradition of the RICE [12] and ANITA [13] detectors, radio sensors have been co-deployed with IceCube strings and more sensors will be deployed this coming season. A larger scale radio detector in a configuration referred to as IceRay is being investigated [14].
• A surface radio component is being investigated to measure the geo-synchrotron signal from cosmic ray air showers in a technique pioneered by LOPES [15]. Such an instrumentation could serve as an extension to IceTop at high energies [7].

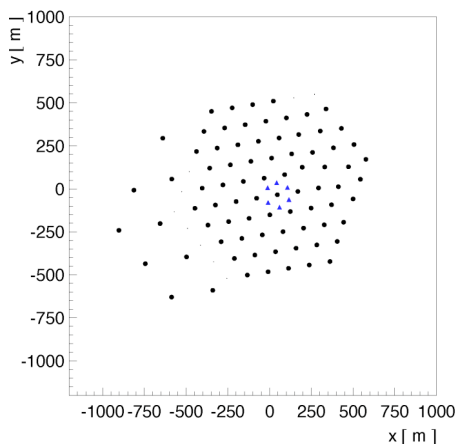

Figure 11: Example of an investigated layout for an end configuration that is optimized for high energies. The 6 strings in the center form the Deep Core, which optimizes IceCube at low energies.

While all of these activities enjoy strong interest inside the IceCube collaboration, it should be noted that the R&D work is still at an early stage. The IceCube collaboration is actively working with outside groups in pursuing the goal of a large scale cosmogenic neutrino telescope at the South Pole.

**Acknowledgments**

We acknowledge the support from the following agencies: U.S. National Science Foundation-Office of Polar Program, U.S. National Science Foundation-Physics Division, University of Wisconsin Alumni Research Foundation, U.S. Department of Energy, and National Energy Research Scientific Computing Center, the Louisiana Optical Network Initiative (LONI) grid computing resources; Swedish Research Council, Swedish Polar Research Secretariat, and Knut and Alice Wallenberg Foundation, Sweden; German Ministry for Education and Research (BMBF), Deutsche Forschungsgemeinschaft (DFG), Germany; Fund for Scientific Research (FNRS-FWO), Flanders Institute to encourage scientific and technological research in industry (IWT), Belgian Federal Science Policy Office (Belspo); the Netherlands Organisation for Scientific Research (NWO); M. Ribordy acknowledges the support of the SNF (Switzerland); A. Kappes and A. Groß acknowledge support by the EU Marie Curie OIF Program.

**References**

[1] E. Waxman and J. Bahcall, Phys. Rev. D 59 (1998) 023002
[2] T. Gaisser, *Cosmic Rays and Particle Physics* Cambridge University Press 1991
[3] M.Ackermann et al., J. Geophys. Res. **111**, D13203 (2006).
[4] J.Ahrens et al., Nucl. Inst. Meth. A **524**, 169 (2004).
[5] M. Ackermann et al., Nucl. Instrum. Meth. A **556**, 169 (2006)
[6] R. Abbasi et al. arXiv:0810.4930, submitted to NIM (2008)
[7] T. Gaisser et al., (IceCube collaboration), Contributions to the ICRC in Merida, arxiv: 0711.0353v1
[8] D. Cowen et al. (IceCube coll.), proceedings of Neutrino 2008 conference, Christchurch, NZ
[9] K. Greisen, Phys. Rev. Lett. **16**, 748 (1966); G. Zatsepin and V.A. Kuz'min, Teor. Fiz. 4, 114 (1966); JETP Lett. 4, 78 (1966).
[10] S. Sarkar, procedings to Neutrino 2008, Christchurch, [arXiv:0811.0375]
[11] J. Vandenbroucke et al., F. Descamps et al., T. Karg et al., B. Semburg et al., all for the IceCube collaboration, D. Tosi et al., these proceedings.
[12] I. Kravchenko et al., Phys.Rev. D **73** (2006) 082002
[13] S. W. Barwick et al., Phys. Rev. Lett. **96** (2006) 171101
[14] H. Landsman et al. (IceCube coll.), J. Kelley et al., these proceedings.
[15] H. Falcke *et al.*, Nature 435, 313 (2005).